# TWO COSMOLOGICAL MODELS,

# THE AGE OF THE UNIVERSE,

# AND "DARK ENERGY"

**Evangelos Chaliasos**

365 Thebes Street

GR-12241 Aegaleo

Athens, Greece

*Summary*

The prevailing cosmological model with the λ-term, in which the space is flat, is studied (§ 1). The corresponding age of the Universe ($t_0$) is calculated (assuming a Hubble constant consistent with the measurements of the Hubble telescope), as well as the deceleration parameter ($q_0$). The latter is found negative, showing an accelerating Universe, but the former is insufficient to account for the actually estimated value. Nevertheless, with more recent values of the parameters involved, this model actually gives a result consistent with the estimated value. But there is a severe defect in the prevailing model, concerning the so-called "dark energy", necessary to be introduced. Then, another model without the λ-term, and based on the time-symmetric theory of the author, is studied (§ 3), after an introduction to this theory (§ 2). In this model the space is open, but the overall space-time is flat. It is not accelerating (it retains a constant rate of expansion). In this model, with $q_0 = 0$, an age of the Universe results, which is consistent again, within the limits of accuracy, with the estimated value. But the great advantage of this model is that it does not require at all the existence of the ambiguous "dark energy".

## 1. Introduction

The purpose of the present paper is to first criticize the prevailing model of the Universe ($\lambda > 0$ & $k = 0$) with regard to the resulting age of the Universe on the one hand and the plausibility of the small cosmological constant implied by the observations on the other hand.

This model, we will study, is "accelerating" ($q_0 < 0$), which can be explained right by the introduction of a positive cosmological constant[*]. That is why we are obliged to introduce a $\lambda > 0$. Besides we take the model to be flat, i.e. $k = 0$, in order to be in conformity to the inflation theory.

We will take for the Hubble constant the value $H_0 = 73 \in (70 \pm 10)$ (km/sec)/Mpc, which is consistent with the observations of the Hubble Telescope [28] in 1999. We will see then that the resulting age of the Universe must be $t_0 \cong 1.25 \times 10^{10}$ yr, instead of the estimated value $t_0 \cong 1.34 \times 10^{10}$ yr [29]. The value found ($1.25 \times 10^{10}$ yr) is of the order of the age of the oldest stars, which has been estimated to be $t \cong 1.2 \times 10^{10}$ yr [29].

For the (matter) density parameter $\Omega_{M0}$ we will take for our model the observed value $\Omega_{M0} \cong 1/3$ [1] (see also map of models in [13]). Then for the contribution of the cosmological constant we will find $\Omega_{\Lambda 0} \cong 2/3$ as follows.

The two equations describing the Universe, with the $\lambda$-term included, are [20]

$$H_0^2 + \frac{kc^2}{S_0^2} - \frac{1}{3}\lambda c^2 = H_0^2 \Omega_0 \tag{1.1}$$

and

---

[*] From (1) & (2) (see below) we have in general, after solving them for $kc^2/S_0^2$ and equating the results, the important relation $\Omega_{M0} = 2q_0 + (2/3)\Lambda(c^2/H_0^2)$ (see [20], p.130, 4.100), which tells us exactly that we can have an accelerating Universe ($q_0 < 0$) if we take $\Lambda$ to be positive "enough".

$$(1-2q_0)H_0^2 + \frac{kc^2}{S_0^2} - \lambda c^2 = 0. \tag{1.2}$$

They give (respectively), for the case k = 0,

$$\Omega_0 = 1 - \frac{1}{3}\lambda \frac{c^2}{H_0^2} \tag{1.3}$$

and

$$q_0 = \frac{1}{2} - \frac{1}{2}\lambda \frac{c^2}{H_0^2}. \tag{1.4}$$

From (1.3), and for $\Omega_{M0} \equiv \Omega_0 = 1/3$, we find $\Omega_{\Lambda 0} \equiv (\Lambda c^2)/(3H_0^2) = 2/3$ (see also for direct measurements [17] and references given there, i.e.: [12], [19], [6], [7], [23], [22]). Then, for this $\Omega_{\Lambda 0}$, we find from (1.4) that $q_0 = -0.5$, that is indeed $q_0 < 0$ and the Universe is accelerating.

But let us start from the beginning. The general relativistic equation of motion is (see, i.e., [17] (p. 1), [27], [24] (p. 252))

$$H^2 \equiv \left(\frac{\dot{a}}{a}\right)^2 = \frac{8\pi G}{3}\rho + \frac{\Lambda c^2}{3} - \frac{kc^2}{a^2}. \tag{1.5}$$

The critical density is

$$\rho_c = \frac{3H^2}{8\pi G}, \tag{1.6}$$

so that

$$\Omega_M \equiv \frac{\rho_M}{\rho_c} = \frac{8\pi G}{3H^2}\rho_M. \tag{1.7}$$

The corresponding to $\rho_M$ densities due to $\Lambda$ & k are

$$\rho_\Lambda = \frac{\Lambda c^2}{8\pi G} \quad (a) \quad \& \quad \rho_k = -\frac{3kc^2}{8\pi G}\frac{1}{a^2} \quad (b), \tag{1.8}$$

so that

$$\Omega_\Lambda \equiv \frac{\rho_\Lambda}{\rho_c} = \frac{\Lambda c^2}{3H^2} \quad (a) \quad \& \quad \Omega_k \equiv \frac{\rho_k}{\rho_c} = -\frac{kc^2}{H^2}\frac{1}{a^2} \quad (b). \tag{1.9}$$

Then (1.5) can be written

$$\rho_M + \rho_\Lambda + \rho_k = \rho_c, \tag{1.10}$$

or, equivalently,

$$\Omega_M + \frac{\Lambda c^2}{3H^2} - \frac{kc^2}{H^2}\frac{1}{a^2} = 1. \tag{1.11}$$

Since

$$\rho_M a^3 = \rho_{M0} a_0^3 = const., \tag{1.12}$$

we can write (1.10) as

$$\frac{\rho_{M0} a_0^3}{a^3} + \frac{\Lambda c^2}{8\pi G} - \frac{3kc^2}{8\pi G}\frac{1}{a^2} = \frac{3\dot{a}^2}{8\pi G a^2}, \tag{1.13}$$

from which

$$\frac{8\pi G}{3}\rho_{M0} a_0^3 \frac{1}{a} + \frac{\Lambda c^2}{3}a^2 - kc^2 = \left(\frac{da}{dt}\right)^2, \tag{1.14}$$

so that

$$dt = \frac{da}{\sqrt{\frac{8\pi G}{3}\rho_{M0} a_0^3 \frac{1}{a} + \frac{\Lambda c^2}{3}a^2 - kc^2}}, \tag{1.15}$$

or, integrating,

$$t = \int \frac{da}{\sqrt{\frac{8\pi G}{3}\rho_{M0} a_0^3 \frac{1}{a} + \frac{\Lambda c^2}{3}a^2 - kc^2}}. \tag{1.16}$$

For <u>k = 0</u>, we have from (1.11) that
$$\Omega_{total} \equiv \Omega_M + \Omega_\Lambda = 1. \tag{1.17}$$

And, since from observations

$$\Omega_{M0} \equiv \frac{8\pi G}{3H_0^2}\rho_{M0} = \frac{1}{3}, \tag{1.18}$$

we obtain from (17)

$$\Omega_{\Lambda 0} \equiv \frac{\Lambda c^2}{3H_0^2} = 1 - \Omega_{M0} = \frac{2}{3}. \tag{1.19}$$

Thus, from the general relation

$$\Omega_{M0} = 2q_0 + \frac{2}{3}\Lambda \frac{c^2}{H_0^2} \tag{1.20}$$

(see[20], p. 130, 4.100)$^\$$, we have

$$\frac{1}{3} = 2q_0 + 2 \cdot \frac{2}{3}, \tag{1.21}$$

from which the result, concerning $q_0$, is

$$q_0 = -0.5 \tag{1.22}$$

Now, from the relation

$$\frac{1}{3} = \Omega_{M0} \equiv \frac{\rho_{M0}}{\rho_{c0}} = \frac{8\pi G}{3H_0^2} \rho_{M0}, \tag{1.23}$$

we obtain

$$\rho_{M0} = \frac{H_0^2}{8\pi G}, \tag{1.24}$$

so that (1.16) gives

$$t_0 = \int_0^{a_0} \frac{da}{a\sqrt{\frac{H_0^2}{3}\left(\frac{a_0}{a}\right)^3 + \frac{2}{3}H_0^2}}, \tag{1.25}$$

or

$$t_0 = \int_0^{a_0} \frac{da}{\frac{H_0}{\sqrt{3}} a\sqrt{\left(\frac{a_0}{a}\right)^3 + 2}}. \tag{1.26}$$

If we set $a/a_0 \equiv x$, then (1.26) becomes

$$t_0 = \frac{\sqrt{3}}{H_0} \int_0^1 \frac{dx}{x\sqrt{\frac{1}{x^3} + 2}} \equiv \frac{\sqrt{3}}{H_0} I. \tag{1.27}$$

We find numerically $I = 0.540331$  Thus $(\sqrt{3})I = 0.935881$, and if we take the value $H_0$ = 73 (km/sec)/Mpc [28], we finally obtain

$$t_0 = \frac{\sqrt{3}I}{H_0} = 1.25 \times 10^{10} \, years, \tag{1.28}$$

which is of the order of the age of the oldest stars $t \cong 1.2 \times 10^{10}$ yr [29].

For comparison, we mention that, for $\Lambda = 0$ & $k = 0$, we have

---

$^\$$ This relation (1.20) can also be written as $q_0 = (1/2)\Omega_{M0} - \Omega_{\Lambda 0}$ and has a general character [30]. It can also be written without the subscript 0 as [3] $q = (1/2)\Omega_M - \Omega_\Lambda$.

$$q_0 = 0.5 \tag{1.29}$$

and

$$H_0 t_0 = 2/3, \tag{1.30}$$

so that

$$t_0 = \frac{2}{3H_0} \cong 0.89 \times 10^{10} \, years, \tag{1.31}$$

which is insufficient to account for the age of the oldest stars $t \cong 1.2 \times 10^{10}$ yr [29].

For values of $\Omega_{M0}$ and $\Omega_{\Lambda 0}$ differrent from those used up to now, the general formula giving $t_0$ becomes

$$t_0 = H_0^{-1} \int_0^1 \frac{dx}{x\sqrt{\Omega_{M0} x^{-3} + \Omega_{\Lambda 0}}}, \tag{1.32}$$

instead of (1.27). For the more recently estimated value $\Omega_{M0} = 0.27 \pm 0.04$ (and therefore $\Omega_{\Lambda 0} = 0.73 \pm 0.04$), we find that the integral above takes on the value 0.9926. If then we use the also more recently estimated value $H_0 = (71 \pm 4)$ (km/sec)/Mpc, consistent again with the value $H_0 = (70 \pm 10)$ (km/sec)/Mpc resulted from the Hubble telescope, we obtain $t_0 = (1.37 \pm 0.02) \times 10^{10}$ years, well above the estimated age of the oldest stars, and in conformity with the value estimated by other methods.[#]

Let us come now to the plausibility of the cosmological constant. Its physical interpretation as vacuum energy density is supported by the existence of the "zero point" energy predicted by quantum mechanics. In this discipline, particle and antiparticle pairs are being created out of the vacuum. Although these particles exist only for a short amount of time before annihilating each other, they do give the vacuum a non-zero energy content (the so-called "dark energy"), *as long as a particle and its antiparticle have the same mass (positive) and therefore positive energies*. In

---

[#] See for example the review article "The New Cosmological Paradigm", By S.Basilakos, in "Hipparchos" (the Hellenic Astronomical Society newsletter), December 2003 (volume 1, issue 13, year 8).

General Relativity all forms of energy should gravitate, including the energy of the vacuum, *hence the cosmological constant.*

But a great problem with the association of the cosmological constant with the above vacuum energy appears when we make even a simple estimate of what this implies for its value. This estimate was done by S.M.Carroll [2] and reproduced in [18]. The result is striking: $\Omega_{\Lambda 0} \sim 10^{120}$ ! In no way can the observed value $\Omega_{\Lambda 0} < 1$ be justified.

However there is an exit out of this big problem. If we assume that for each created particle, with mass say m (positive), its corresponding antiparticle has the *opposite* mass, i.e. –m (*negative*), so that, including their (rest) masses, their energies also cancel each other, the net effect will be zero, resulting in $\Omega_\Lambda = 0$. We will see in what follows that the small value $\Omega_\Lambda < 1$, and the associated "dark energy", is in fact not necessary. This is the case since, because of the opposite masses of a particle and its corresponding antiparticle mentioned above, the whole situation changes *radically*, as we will see in what follows.

### 2. The time-symmetric theory in brief

*A. Generalities and definitions.*

*Motion backwards in time and the principle of anticausality.*

It is well known that the elementary interval (the line element in the Minkowski Geometry) is given in the Special Theory of Relativity by the relation

$$ds^2 = c^2 dt^2 - dl^2, \tag{2.1}$$

where c is the speed of light (in vacuum), t is the coordinate time, and dl is the elementary length. Extracting $c^2 dt^2$ as a common factor, we have

$$ds^2 = c^2 dt^2 \left(1 - \frac{v^2}{c^2}\right) \tag{2.2}$$

where v is the coordinate velocity of a point tracing ds, which is given by the formula

$$v = dl/dt. \tag{2.3}$$

But if we want to introduce the *proper time* τ, then we have that

$$ds^2 = c^2 d\tau^2. \tag{2.4}$$

Thus, from the relations (2.2) and (2.4) we take

$$d\tau^2 = dt^2 \left(1 - \frac{v^2}{c^2}\right) \tag{2.5}$$

and, extracting the square root, we end in the relation

$$d\tau = \pm dt \sqrt{1 - \frac{v^2}{c^2}}, \tag{2.6}$$

where we consider that $|v| \leq c$ in order for the proper time to be real. It must be noted here that we usually consider only the *plus sign (+)* in the relation (2.6). But, as we will see in what follows, it is *necessary* to also consider the *minus sign (-)*, in order to describe and study the motion *backwards in time*. Of course, if we want to go to Classical Mechanics, we have to assume that v/c<<1, so that (2.6) leads us to

$$d\tau = \pm dt. \tag{2.7}$$

That is Relativity is not necessary to study the motion backward in time.

Thus, concerning normal particles, that is those moving forward in time, it is necessary to consider the *plus sign (+)* in the relations (2.6) and (2.7), while, concerning particles moving (perhaps) backwards in time, we have to consider the *minus sign (-)* in the relations (2.6) and (2.7). We call *retrons* these latter (assumed) particles and I claim that such particles actually exist, as it will be seen in what follows. But the direction of the motion (forward or backward in time) is directly related with the *principle of causality,* according to which the *cause* <u>always</u> preceds *temporarily* (in respect to the *coordinate* time) the *effect.* In fact, if we consider two events along the orbit of a particle and we parametrize the curve of the orbit with the proper time τ as a parameter, and if we assume that τ(A)<τ(B), then we may consider A as an emission of the particle and B as an absorption, where the emission *logically* (in respect to *proper* time) precedes the absorption. But, because the emission of the particle may be considered as *cause,* with the absorption as the corresponding *effect,* we can say that <u>always</u> the cause *logically* (in respect to *proper* time) precedes the effect. We distinguish two cases: 1) if it is about a normal particle, and because we have to take the *plus sign (+)* in the relations (2.6) and (2.7), the cause A (emission) will precede the effect B (absorption) also *temporarily* (in respect to the *coordinate* time) that is the principle of causality will hold, as long as it occurs in *every* (inertial) reference frame (a fact which is secured by $|v| \leq c$). And 2) if it is about a retron, and because we have to take the *minus sign (-)* in the relations (2.6) and (2.7) in this case, the cause A (emission) will *temporarily* <u>follow</u> the effect B (absorption), in other words the principle of causality will be violated. If this happens in *every* (inertial) reference frame (a fact which also occurs for $|v| \leq c$), we may say that a *principle of anticausality* holds for retrons.

The two principles exposed above (those of causality and of anticausality) are related, if we consider the corresponding *world lines* in the Minkowski plane, the first one with the motion from the origin of coordinates O *upwards* and *inside* the light cone of O (*absolute future*) concerning normal particles, and the second one with motion from the origin of coordinates O *downwards* and *inside* the light cone of O (*absolute past*) concerning retrons. In both cases we have a *timelike* world line, because

$$ds^2 \geq 0, \tag{2.8}$$

or

$$(ct - x)(ct + x) \geq 0 \tag{2.9}$$

in one (spatial) dimension, for simplicity. We observe that this happens when either

$$ct - x \geq 0 \quad \& \quad ct + x \geq 0 \tag{2.10}$$

(the principle of *causality* being satisfied and the motion being *forward* in time), or

$$ct - x \leq 0 \quad \& \quad ct + x \leq 0 \tag{2.11}$$

(the principle of *anticausality* being satisfied and the motion being *backward* in time). Because the space-time elementary interval $ds^2$ remains invariant when we go from one reference frame to another, it follows that both principles of causality and of anticausality have *absolute* character. I propose the generalization of the "principle of causality" to a *principle of generalized causality*, which will include exclusively the principle of causality (followed by normal particles) on the one hand, and the principle of anticausality (followed by retrons) on the other hand. It is evident from the above, that the principle of generalized causality remains invariant (it continues to hold) if we go from one reference frame to another, so following the Einstein Principle of (Special) Relativity.

We have to observe here that the assumed particles called *tachyons* (that is particles having velocities > c) *do not follow* the Einstein Principle of (Special) Relativity concerning causality, and this is the case because their world line is *spacelike,* and therefore the cause (emission) precedes the effect (absorption) in some reference frames, while the cause follows the effect in other ones. Thus, tachyons do not comply to either the principle of causality (in all reference frames), or the principle of anticausality (in all reference frames), that is the principle of generalized causality is violated by them. Thus, if we consider the principle of generalized causality (or each of its constituent principles of causality and of anticausality) as a physical law (which has to necessarily follow the Principle of Relativity), tachyons cannot exist theoretically.

B. *Negative mass and opposite electric charge.*

It is well known that the motion of a normal particle is described (in the most general way) by the Principle of Least Action. According to this principle a function L (the *Lagrangian*) is associated with the particle, such that the time integral of the Lagrangian along a virtual path with known ends (that is the *action* S) from the *original* moment of time $t_A$ to a *final* moment of time $t_B$

$$S = \int_{t_A}^{t_B} L dt \qquad (2.12)$$

becomes minimum for the actual motion of the particle. Of course $t_A < t_B$ for normal particles. Then, if the particle is free (that is it is not subject to forces), it can be shown [14] that the Lagrangian is given by the relation

$$L = \frac{1}{2} mv^2, \qquad (2.13)$$

where v is the velocity of the particle and m is a factor of proportionality called the *mass* of the particle. In addition, in order for the action S, given by (2.12), to have a minimum, it is necessary the mass m in the Lagrangian (2.13) to be *positive*. This is the case because, if it was negative, we could make L (cf. relation (2.13)) negative and as small as we wanted (increasing the velocity v), so that S (cf. (2.12)) could not have a minimum.

Now in the case of a retron following the same world line but in the *opposite* direction, L will change in general and from L it will become L´, such that the action S is now given by the formula

$$S = \int_{t_B}^{t_A} L' dt, \tag{2.14}$$

with the limits of integration $t_B$ and $t_A$ as before but in the opposite ordering. We want the new Lagrangian $L'$ to retain the same form, that is to be given again by a relation of the form

$$L' = \frac{1}{2} m' v^2, \tag{2.15}$$

where the new mass m´ may differ from m. And because again $t_B > t_A$, the same reasoning as before (in order for the action S to have a minimum) gives us
$$L' = -L, \tag{2.16}$$

and therefore the mass of the retron will be *negative,* and especially
$$m' = -m. \tag{2.17}$$

For a charged, with charge e, normal particle of mass m > 0, which is located in an electromagnetic field, produced by an electromagnetic potential ($\varphi$, **A**), the Lagrangian is given [15] by the relation

$$L = -mc^2 \sqrt{1 - \frac{v^2}{c^2}} + \frac{e}{c} \vec{A} \cdot \vec{v} - e\phi. \tag{2.18}$$

Passing to the corresponding retron, we will similarly have

$$L' = -m'c^2 \sqrt{1 - \frac{v^2}{c^2}} + \frac{e'}{c} \vec{A} \cdot \vec{v} - e'\phi. \tag{2.19}$$

And because of (2.16) and (2.17), we will obviously have for this retron
$$e' = -e. \tag{2.20}$$

In other words the original (normal) particle changes sign in its charge in order to give us the corresponding retron.

Also, we have to observe that for the same normal particle the equation of motion is given by the Lorentz force

$$mc \frac{du^i}{ds} = \frac{e}{c} F^{ik} u_k, \tag{2.21}$$

where $u^i$ is its four-velocity and $F^{ik}$ is the electromagnetic field tensor. For the corresponding *antiparticle* now, we similarly have

$$mc\frac{du^i}{ds} = \frac{e'}{c}F^{ik}u_k, \tag{2.22}$$

with e´ given by (2.20). Feynman [10] first considered antiparticles as "normal" particles, which however <u>move backwards in time</u>, so that we have to write for them, instead of (2.22), the relation

$$mc\frac{du^i}{ds'} = \frac{e}{c}F^{ik}u_k, \tag{2.23}$$

with

$$ds' = -ds. \tag{2.24}$$

As we see, the relations (2.22) and (2.23) are equivalent. But a particle moving backwards in time cannot be a normal one, on the contrary being *by definition* a retron. Thus, in Feynman´s antiparticle, we have to take mass and charge given by the relations (2.17) and (2.20). Then (2.23) becomes <u>equivalently</u>

$$m'c\frac{du^i}{ds'} = \frac{e'}{c}F^{ik}u_k. \tag{2.25}$$

It is a curious fact why Feynman was not led from (2.23) to (2.25), although he knew that antiparticles have *negative energy*. In fact, Feynman knew that, from the relation

$$E^2/c^2 = p^2 + m^2c^2 \tag{2.26}$$

of Special Relativity, where E is the energy and p the momentum of the particle, we take two values for the energy, the values

$$E = \pm c\sqrt{p^2 + m^2c^2}, \tag{2.27}$$

where the plus sign (+) corresponds to a normal particle, and the minus sign (-) corresponds to its antiparticle. But then, from the relation

$$E = \frac{mc^2}{\sqrt{1-\frac{v^2}{c^2}}}, \tag{2.28}$$

it is necessary but a little "courage" to introduce negative masses, and, thus, <u>to identify antiparticles with retrons!</u>

On the other hand the claim, that antiparticles having positive masses is an "experimental fact", is wrong, as it will be seen in what follows. Moreover since, as we have already seen above (see the paragraph before the last one of the previous section), positive masses of antiparticles would lead to an enormously large value of the cosmological constant, which is not confirmed by experiment! Thus, it necessarily follows that antiparticles must have <u>negative</u> masses, so that we have to <u>identify</u> them with retrons, moving backwards in time.

Concerning the ratio charge/mass in positrons (for example), as compared with the same ratio for electrons, I have to admit that it is given in principle by e´/m´, where e´ = -e and m´ = -m, that is e´/m´ = e/m, as it can be seen from eqn. (2.25), the equation of motion for antiparticles. But we have to observe that ds´ = -ds enters in this equation rather than ds. Nevertheless, when we observe a positron, because our time runs opposite to the positron´s proper time, we will see the positron as if it were moving forward in time (by simply observing it successively at earlier and earlier instants of its proper time). In other words we can by no means be aware of its movement backwards in time, and we thus attribute ds rather than ds´ to its space-time elementary interval, which of course we enter into eqn. (2.25). But, in order for the equation to remain equivalent to itself, we are oblidged to substitute m´ = -m by m in eqn. (2.25). In this way, the ratio of charge/mass for positrons appears to be e´/m = -e/m, consistent with eqn. (2.22), obviouslydifferent from eqn. (2.21) appropriate for electrons.

Concerning the validity of the law F = ma in respect to positrons, we have to observe the following. At first, in respect to electrons, we have F = ma, with F > 0 and a > 0 (say), and m > 0. In respect to positrons, we have both F and m to change sign, that is F < 0 and m < 0. Thus we have to take a > 0 again, in order for F = ma to hold.

But, since we are unbelieving in taking m < 0 and we insist in taking m > 0 again, we have to take a < 0 in order for F to remain unchanged. The latter case corresponds to motion forward in time (m > 0). Otherwise we have the case of motion backwards in time. The two cases are taken from each other by simultaneously changing the sign in both a and m. I will explain it better. In the meanwhile, it has to be stressed again that we are unable to "see" the motion backwards in time. We will again see the motion as if it were taking place forward in time, and this is the case because <u>our</u> time elapses "forward" (opposite to the antiparticle´s proper time), so that we see the positron at earlier and earlier instants of its proper time. This is another reason why we are inclined to take a < 0, that is opposite to that of electrons (with the result to be oblidged to take again m > 0).

In an electron-positron pair creation, the key point is that in fact the antiparticle does experience the same acceleration as the particle, if the former is thought of as moving on the same world line but backwards in time having negative mass. This is the case for the following reasons. First, the velocity $v = dx/d\tau$ of the antiparticle, with $\tau$ the proper time, changes sign for motion backwards in time and on the same world line, for $dx \to -dx$ and $d\tau$ (= -dt now with dt < 0) remains the same. Then $dv \to -dv$ for the antiparticle as we go to motion backwards in time on the same world line, with $d\tau$ as we said being the same in sign as in motion forward in time. Thus $a = dv/d\tau$ will change sign for the antiparticle if we go from motion forward in time on the well known world line to motion backwards in time on the same world line, that is a will be the same as in the case of the particle, since a for the antiparticle, if it is thought of as moving forward in time, is evidently the opposite of a for the particle.[(&)]

---

[(&)] Concerning the energy threshold of $2mc^2$ in an electron-positron pair creation, in my theory, the pair creation (completely symmetric to the electron-positron annihilation) is given by the reaction $\gamma + \beta^+ \to \beta^- + \gamma^*$, where $\beta^+$ and $\gamma^*$ are the antiparticles of $\beta^-$ and $\gamma$ respectively, the former moving backwards in

*C. Space-time transformation, antialgebra and MCPT symmetry.*

*Antiworld.*

Let us perform the space-time transformation of *time reversal* t → t´ = -t. At first glance this transformation must not influence the space coordinates. But this is not the case. Let us take as an example the x coordinate. Evidently the direction of the x-axis is determined by the direction of motion of a moving particle which traces it. But, with the time reversal mentioned above, the same moving particle appears to move in the *opposite* direction! And the same thing happens for the other two space axes, too. In other words time reversal *induces* a change of direction for *all* axes. Thus, the time reversal t → t´ = -t *necessarily* results in the space inversion **x** → **x**´ = -**x** also.

This fact has the following consequence. The Lorentz transformations must be written

$$\tau = \varepsilon \cdot \beta \left(t - vx/c^2\right) \tag{2.29a}$$
$$\xi = \varepsilon \cdot \beta \left(x - vt\right) \tag{2.29b}$$
$$\eta = \varepsilon \cdot y \tag{2.29c}$$
$$\zeta = \varepsilon \cdot z, \tag{2.29d}$$

where

$$\beta = \frac{1}{\sqrt{1 - \frac{v^2}{c^2}}}, \tag{2.30}$$

and
$$\varepsilon = \pm 1, \tag{2.31}$$

with (t, x, y, z) the old coordinates and (τ, ξ, η, ζ) the new ones. This results from the derivation of the transformations by Einstein himself [9]. In fact, Einstein arrives

---

time. Accordingly, if we take in mind that energy is negative for motion backwards in time, the energy balance is hν/2 – mc² = mc² – hν/2, or hν = 2mc². We thus have again an energy threshold of 2mc².

originally to the transformations (2.29), where, instead of ε, a function φ(v) appears. Then, he finds, concerning φ(v), the relations

$$\varphi(v)\ \varphi(-v) = 1 \tag{2.32a}$$

and

$$\varphi(v) = \varphi(-v). \tag{2.32b}$$

He then concludes that φ(v) = 1, *disregarding the equally possible case φ(v) = -1*. Thus, if we consider this case as equally possible, we see at once that (for v = 0) *time reversal necessarily results in space inversion as well!*

Let us now have the (inertial) reference frames S and S´, where the second one results from the first one after a combined time reversal and space inversion. Let us also consider an observer of S´ who performs binary operations, in particular multiplications, with the coordinates of his frame. He will undoubtedly apply the rules

$$+\cdot+ = +,\ +\cdot- = -,\ -\cdot+ = -,\ -\cdot- = + \tag{2.33}$$

The question is how will an observer of S interpret the above results. Let us take the binary operation + . - = - of S´, as an example. Obviously the observer of S must perform the transformations

$$+ \to -\ \ \&\ \ - \to + \tag{2.34}$$

But then, for this binary operation of S´, the result in S will be - . + = + ! It is thus obvious that, to the binary operation of multiplication (.), we have to correspond *another* binary operation, say *, which we call *antimultiplication* and is given symbolically by the rules

$$+*+ = -,\ +*- = +,\ -*+ = +,\ -*- = - \tag{2.35}$$

We call *antialgebra* the algebra based on this new binary operation *. It is thus evident that, in order to perform binary operations in S (multiplications [and divisions]) referring to coordinates of S´, we have to *first* change the signs in these coordinates, and *second* to use the rules of antialgebra!

Thus, we saw that time reversal has as a consequence the complete set of transformations

$$\left.\begin{aligned} m &\to m' = -m & \text{(a)} \\ e &\to e' = -e & \text{(b)} \\ t &\to t' = -t & \text{(c)} \\ \bar{x} &\to \bar{x}' = -\bar{x} & \text{(d)} \end{aligned}\right\} \quad (2.36)$$

If we call M the transformation (2.36a), C the transformation (2.36b), T the transformation (2.36c), and P the transformation (2.36d), it then results that the composite transformation MCPT leaves the whole physical reality invariant, in other words we cannot distinguish between the image and the original.

It is a remarkable fact that the physical quantities corresponding to the transformations M, C, P, T are right the fundamental quantities of the system of units MKSA. This fact, in combination with the fact that *all* physical quantities are expressed as functions of the fundamental quantities of MKSA, leads us to the conclusion that, inverting the signs of the fundamental quantities and applying of course antialgebra concerning quantities of S′ estimated in S, *every* physical quantity changes sign when it is transformed from S to S′ and *vice versa,* in other words the *arbitrary* physical quantity X of S is transformed when we go to S′ as

$$X \to X' = -X. \qquad (2.37)$$

The time symmetry, if antiparticles are interpreted as retrons, leads necessarily to a complete symmetry of matter and antimatter in the Universe. If in addition we assume that the Universe is *homogeneous* and *isotropic* at large scale, as it is really observed, matter must be *uniformly* mixed with antimatter, of course at large scale (that is the fluctuations giving rise to matter and antimatter structures occupy different regions of the Universe, in this way being prevented from annihilation). If then we call *World* the totality of matter, the totality of antimatter must be called *Antiworld.* We will see in the next section that phenomena have already been discovered which, if they cannot be called proofs, they certainly constitute indications for the real existence of this Antiworld! Note that, if the Antiworld really exists, then *all* physical phenomena in it must evolve *reversely* in time!

### 3. A new cosmological model

For the Robertson-Walker metric the Einstein equations, with the $\Lambda$-term included, are [24]

$$\frac{\dot{R}^2}{R^2} - \frac{8\pi\, G\rho}{3} = -\frac{kc^2}{R^2} + \frac{\Lambda c^2}{3} \qquad (3.1)$$

and

$$\frac{\dot{R}^2}{R^2} + 2\frac{\ddot{R}}{R} + \frac{8\pi\, Gp}{c^2} = -\frac{kc^2}{R^2} + \Lambda c^2. \qquad (3.2)$$

If we restrict ourselves, mainly for theoretical reasons [15], to their classical form, i.e. without the $\Lambda$-term (setting $\Lambda = 0$), they reduce to the Friedmann equations, which are therefore

$$\frac{\dot{R}^2}{R^2} - \frac{8\pi\, G\rho}{3} = -\frac{kc^2}{R^2} \qquad (3.3)$$

and

$$\frac{\dot{R}^2}{R^2} + 2\frac{\ddot{R}}{R} + \frac{8\pi\, Gp}{c^2} = -\frac{kc^2}{R^2}. \qquad (3.4)$$

Let us apply them to a Universe consisting of the World and the Antiworld. In order for the Universe to retain homogeneity and isotropy at large scale, it is necessary for the World and the Antiworld to be mixed at the same scale. Thus we will consider the Universe as made out of Islands of matter and Antiislands of antimatter, equal in number and consisting of the same amount of matter and antimatter respectively (on the average). These are supposed to be more likely clusters (and anticlusters) of galaxies (and antigalaxies), which thus constitute the "particles" (and the "antiparticles") of the cosmic fluid.

Since the mass of the "antiparticles" is opposite (negative) of the mass of the "particles" according to the theory proposed in the previous section, they must cancel each other, so that we have to take for the cosmic fluid ρ = 0. Also, since obviously the "antiparticles" have momenta of opposite magnitudes (negative) as compared with the "particles", having positive ones, it is clear that they will exert opposite pressures. And of course it is evident that radiation and "antiradiation" will exert opposite pressures as well. Thus the net effect is that the overall total pressure of the cosmic fluid must be p = 0.

The result is that our equations (3.3) & (3.4) become

$$\frac{\dot{R}^2}{R^2} + \frac{kc^2}{R^2} = 0 \tag{3.5}$$

and

$$2\frac{\ddot{R}}{R} + \left(\frac{\dot{R}^2}{R^2} + \frac{kc^2}{R^2}\right) = 0. \tag{3.6}$$

The second one, (3.6), of course reduces to
$$\ddot{R}/R = 0, \tag{3.7}$$

because of the first one, (3.5).

In order to solve them for R, we have to observe at first glance, because of (3.5), that k must be negative, that is obviously k = -1, and we have to deal with hyperbolic space slices. From (3.5) then we have
$$\dot{R}^2 = c^2, \tag{3.8}$$

so that, integrating, we are left with the very simple result
$$R = \pm ct. \tag{3.9}$$

If we go to (3.7) now,

$$\ddot{R} = 0, \tag{3.10}$$

we simply observe that it is automatically satisfied (it becomes an identity) upon substituting R with the solution found, (3.9).

Thus our Universe expands uniformly (i.e. without either deceleration or acceleration), since

$$q \equiv -\frac{\ddot{R}R}{\dot{R}^2} = 0. \tag{3.11}$$

We also have from (3.9) that

$$H \equiv \frac{\dot{R}}{R} = \frac{1}{t}. \tag{3.12}$$

From (3.12), we see then that the age of the Universe is simply given by

$$t_0 = \frac{1}{H_0}. \tag{3.13}$$

Substituting $H_0 = 73$ (km/sec)/Mpc [28] in (3.13), we take the result $t_0 = 1.34 \times 10^{10}$ yr, in complete accordance with the estimations [29], and larger than the estimated age of the oldest stars $t = 1.2 \times 10^{10}$ yr [29]. Using the more recently estimated value $H_0 = (71 \pm 4)$ (km/sec)/Mpc, also consistent with the value $H_0 = (70 \pm 10)$ (km/sec)/Mpc resulted from the Hubble telescope, we find $t_0 = (1.38 \pm 0.08)$ years, a value in accordance with the value $t_0 = (1.37 \pm 0.02)$ years resulted from the prevailing model.

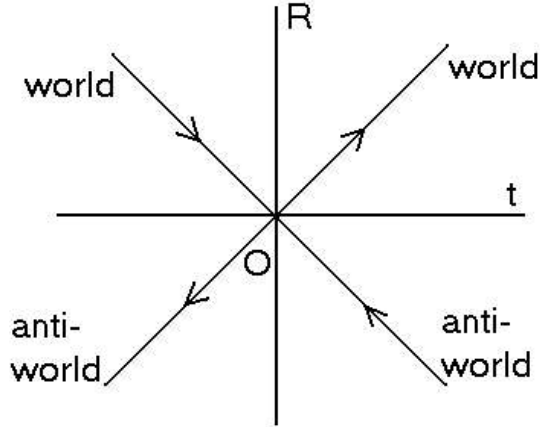

Fig.1

Fig.1: The solution R = ±ct, (3.9), for the Universe. Note that above the t-axis the proper time (shown by arrows) elapses as the coordinate time (world) because R > 0 there. The opposite (antiworld) happens below the t-axis.

Despite the fact that k = -1, we must emphasize here that the overall space-time is flat in this model. This can be seen from the fact that our metric now is, because of (3.9),

$$ds^2 = c^2 dt^2 - c^2 t^2 \{d\chi^2 + \sinh^2\chi\, (d\vartheta^2 + \sin^2\vartheta\, d\varphi^2)\}, \qquad (3.14)$$

which, by the substitution
$$r = ct \sinh \chi \quad (a), \quad \tau = t \cosh \chi \quad (b), \qquad (3.15)$$

can be transformed to the form
$$ds^2 = c^2 d\tau^2 - dr^2 - r^2 (d\vartheta^2 + \sin^2\vartheta\, d\varphi^2), \qquad (3.16)$$

that is to Galilean form [15].

We can see that our model resembles the Milne universe. And also that the singularity of the Friedman models defining the beginning of time disappears because of (3.16). In other words the "origin" O (see Fig.1) of time is no longer a singularity.

And, from the form of Fig.1, O can be described as a "big collision" rather than the "big-bang".

Let us now come to the "necessity" for the introduction of the cosmological constant. $\Lambda$ was introduced in order to explain the data from some very distant supernovae ([11], [13]) which suggested that the Universe is not decelerating but instead it can be even accelerating. Thus, the introduction of $\Lambda$ resulted in a universal force of cosmic repulsion, which was opposed to the attraction due to the material content of the Universe. But it is not necessary to do this. In fact, in our model, while matter attracts matter, as well as antimatter attracts antimatter, according to Newton´s law

$$F = G\frac{m_1 m_2}{r^2} > 0 \qquad (3.17)$$

and, respectively,

$$F = G\frac{(-m_1)(-m_2)}{r^2} > 0, \qquad (3.18)$$

matter repels antimatter, as well as antimatter repels matter, according to (the same) Newton´s law

$$F = G\frac{m_1(-m_2)}{r^2} < 0 \qquad (3.19)$$

and, respectively,

$$F = G\frac{(-m_1)m_2}{r^2} < 0. \qquad (3.20)$$

Thus, we can have the necessary repulsion without having to introduce the exotic cosmological constant $\Lambda$.

We will see that the redshift $|z'|$ corresponding to an antimatterial object is less than the redshift $z$ corresponding to an equally distant material object, if the observer lies on the straight line joining them. Of course the source of $z'$ will belong to the antiworld, so that its light emitted will move backwards in time, with the result to reach us coming from the future. But, before proceeding to show how can $|z'| < z$ in the particular case in our model of the Universe, it is important to answer to two questions that can be risen, even from the beginning.

The first question is that, since light moving backwards in time has negative energy, say –E, we should take its absorption by the telescope as emission. This is in fact the case, because if the light is absorbed in a time interval -t, which of course is negative (since the light moves *backwards* in time), then the flux $\Phi = (-E)/(-t)$ incident on the telescope will be *negative* if we use antialgebra, as we have to do since the light is coming from the Antiworld. Thus, we will observe *emission* rather than absorption (of course if the emulsion used in the telescope can *emit*).

The second question is that, since we will have greater size of the expanding Universe (with scale factor a) in the future, we should observe blueshift instead of redshift. This is again the case, because if

$$z = \frac{\omega}{\omega_0} - 1 = \frac{a_0}{a} - 1 > 0 \qquad (3.21)$$

for light coming from the past (i.e. moving forward in time), then for light coming from the future (i.e. moving backwards in time) we should attribute to an anti-observer as spectral shift the opposite of that given by (3.21), that is

$$\tilde{z}' = 1 - \frac{\tilde{\omega}'}{\tilde{\omega}_0} = 1 - \frac{\tilde{a}_0}{\tilde{a}'} > 0, \qquad (3.22)$$

i.e. again a *redshift*. Note that if we were also belonging to the antiworld, then we would observe a blue shift, given by

$$z' = \frac{\omega'}{\omega_0} - 1 = \frac{a_0}{a'} - 1 < 0 \qquad (3.23)$$

(i.e. a relation analogous to (3.21). Then an observer of the world should change all signs in (3.22), with the result to have a relation similar to (3.23), and thus to observe again a *blueshift* z´ given right by (3.23).

Coming now to the assertion made above, that $|z'| < z$ for equally distant sources belonging to the antiworld the first and to the world the second, and in the particular case mentioned above, we have to state first that we mean "light distance" [4], that is given by $d' = c(t' - t_0)$ and $d = c(t_0 - t)$ respectively. Then because of (3.23) and (3.21) we must show that

$$1 - \frac{a_0}{a'} < \frac{a_0}{a} - 1, \tag{3.24}$$

or

$$\frac{a' - a_0}{a'} < \frac{a_0 - a}{a}. \tag{3.25}$$

But, because of (3.9), the numerators in (3.25) are exactly the light distances, which are taken to be equal. Thus, from $a' > (a_0 >) a$ the relation (3.25) becomes obvious, QED. Of course "luminosity distances" D, defined by the requirement that the surface of a sphere of radius D be equal to $4\pi D^2$, are used rather than light distances d. But, because the space time is flat, they are the same as above. The general result is that the relation $|z'| < z$ is false, as it will be seen from what follow (see Appendix A).

Another useful, in what follows, formula comes from the relation [15]

$$ad\eta = cdt, \tag{3.26}$$

or, in our case,
$$\cancel{c}td\eta = \cancel{c}dt. \tag{3.27}$$

Integrating this relation, (3.27), we find simply
$$\eta = \ln t. \tag{3.28}$$

Let us now find the distance-redshift relation for both the world and the antiworld, using luminosity distances in order for them to be easily converted to apparent magnitudes. The result will be to take finally the magnitude-redshift relation, which can be directly compared with the observations. And we will work, without approximations, in full generality, since the data extend even to z = 1 today. Let us find it first for the world. We will start from the relation [25]

$$D = a(\eta_0)\Sigma(\chi)(1+z), \qquad (3.29)$$

which for k = -1 becomes

$$D = a(\eta_0)\sinh\chi \,(1+z). \qquad (3.30)$$

Because of (3.9), and since [15]

$$\chi = \eta_0 - \eta, \qquad (3.31)$$

which becomes, because of (3.28),

$$\chi = \ln\frac{t_0}{t}, \qquad (3.32)$$

the relation (3.30) gives

$$D = ct_0 \sinh\left(\ln\frac{t_0}{t}\right)(1+z), \qquad (3.33)$$

which results in

$$D = \frac{1}{2}ct_0\left\{\frac{t_0}{t} - \frac{t}{t_0}\right\}(1+z), \qquad (3.34)$$

or

$$D = \frac{1}{2}a_0\left\{\frac{a_0}{a} - \frac{a}{a_0}\right\}(1+z). \qquad (3.35)$$

Doing the calculations and using (3.21), we obtain

$$D = \frac{1}{2}ct_0 \cdot z\left(1 + \frac{a}{a_0}\right)(1+z) \qquad (3.36)$$

or, because of (3.13), and (3.21) again,

$$D = \frac{cz}{2H_0}\left(1 + \frac{1}{1+z}\right)(1+z), \qquad (3.37)$$

which gives the final result

$$D = \frac{cz}{2H_0}(2+z). \qquad (3.38)$$

In a similar fashion, we find for

$$D' = a(\eta_0)\Sigma(\chi')(1+z'), \tag{3.39}$$

if we use (3.23) instead of (3.21) and taking in mind that [15]
$$\chi' = \eta' - \eta_0, \tag{3.39'}$$

that finally it is given by
$$D' = -\frac{cz'}{2H_0}(2+z'). \tag{3.40}$$

We will see an alternative way of extracting (3.38) and (3.40) in Appendix B.

Using the astronomical photometric relations
$$m - M = 5\log(D/pc) - 5, \tag{3.41}$$

and respectively
$$m' - M = 5\log(D'/pc) - 5, \tag{3.42}$$

we can find immediately from (3.38) & (3.40) respectively the corresponding, exact, magnitude-redshift relations. From the approximate relation [4],

$$\pm\frac{H_0 D}{c} \cong z - \frac{1}{2}(q_0 - 1)z^2, \tag{3.43}$$

which becomes linear for $q_0 = 1$, we see that both (3.38) and (3.40) correspond to $q_0 = 0$. Note that, from the relations (3.21) and (3.23), z extends from 0 to $+\infty$ while z´ extends from 0 to -1. See Fig.2 for the variety of curves corresponding to $q_0$ = 0, 1, 2 ($\Lambda$ = 0). Note also the excellent agreement of the $(\Omega_M, \Omega_\Lambda) = (0,0)$ curve of Fig.3 to the data. This curve corresponds to $(q_0, \Lambda) = (0, 0)$ of our model.

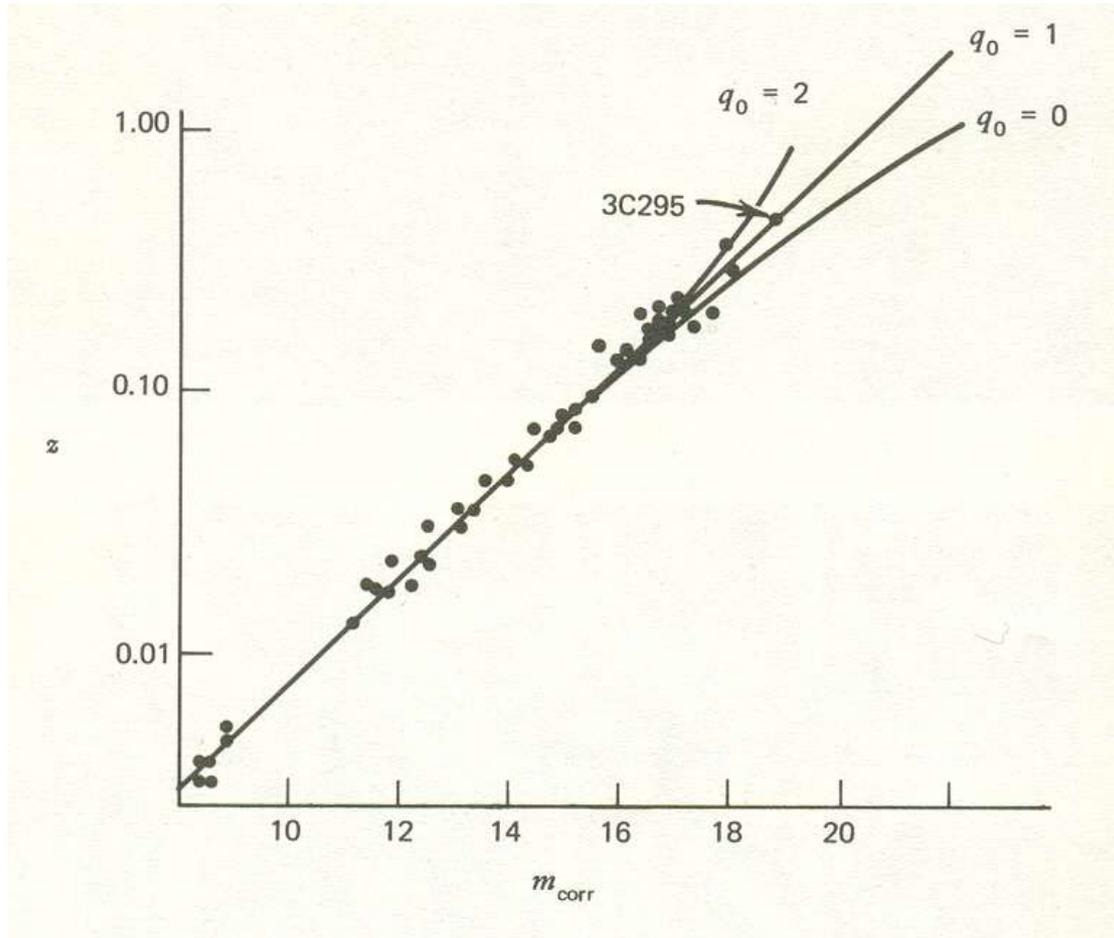

**Fig.2** Red shifts versus corrected apparent magnitudes for 42 first-ranked cluster galaxies. The curves represent fits of Eqn. (3.43) to the data, for $q_0$ = 0, 1, 2. The figure is taken from [31].

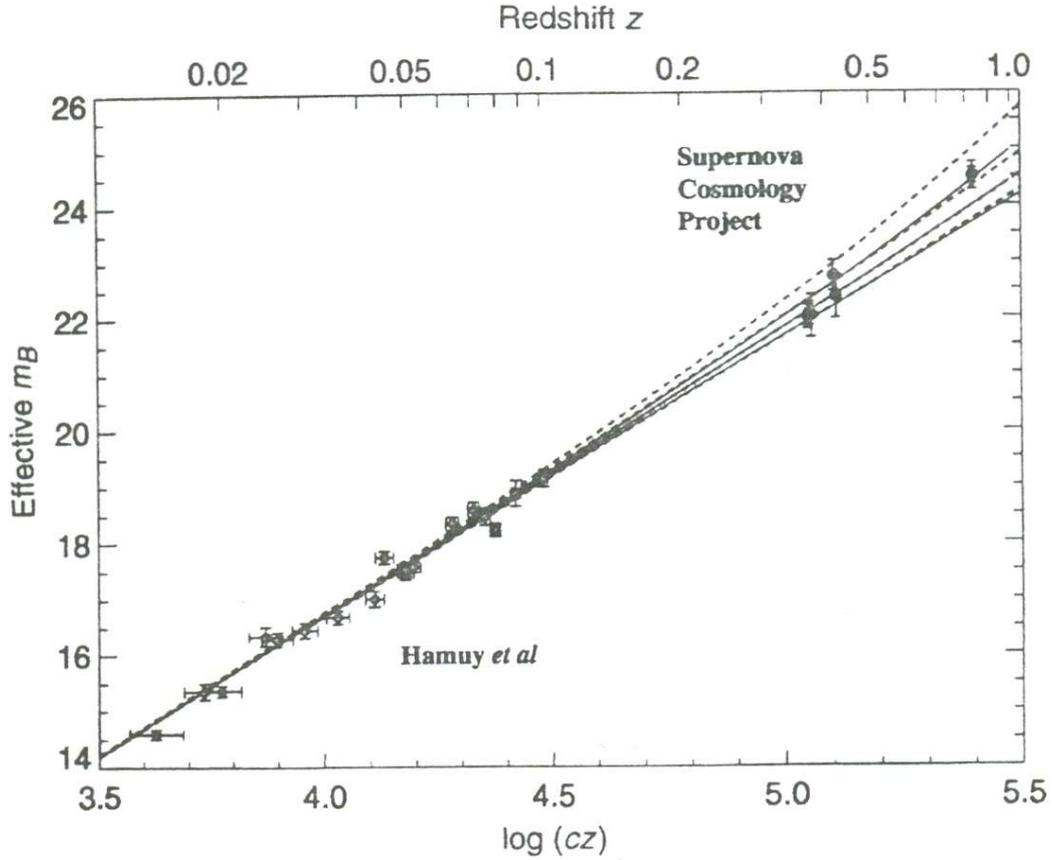

**Fig.3** SN1997ap at z = 0.83 plotted on the Hubble diagram. Also plotted are the 5 of the first 7 high-redshift supernovae that could be width-luminosity corrected, and the 18 of the lower-redshift supernovae from the Calan/Tololo Supernovae Survey that were observed earlier than 5 d after maximum light. Magnitudes have been K-corrected, and also corrected for the width-luminosity relation. The solid curves are theoretical $m_B$ curves for $(\Omega_M, \Omega_\Lambda)$ = (0, 0) on top, (1, 0) in the middle and (2, 0) on bottom. The dotted curves are for the flat-universe case, with $(\Omega_M, \Omega_\Lambda)$ = (0, 1) on top, (0.5, 0.5), (1, 0), and (1.5, -0.5) on bottom. The figure is taken from [21].

Ending this work, we have to mention two ways by which one could wonder how can we distinguish between the light of a source if it comes from the future (moving backwards in time) (case b) and the light of the same source if it comes from the past (moving forward in time) (case a). In fact, there are two ways which enables us to distinguish between the two cases.

First, by observing any *asymmetric* phenomenon taking place, for example the light curve of a supernova (see fig. 4). Namely, since a normal supernova (belonging to a normal galaxy) gives the light curve (a), an antisupernova (belonging to an antigalaxy) must give the light curve (b).

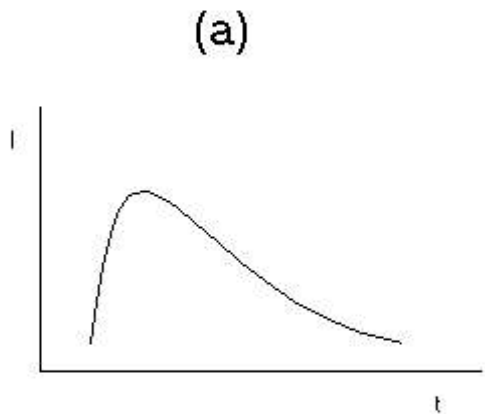
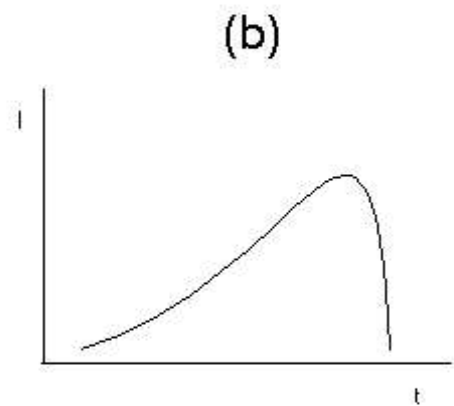

Fig. 1: light curve of a supernova

Fig. 2: light curve of an anti-supernova

Fig. 4. Light curves

The second way is right to examine the light´s aberration. It is obvious that the aberration occurs in opposite directions in these two cases. Thus, examining the source´s aberration ellipse (due to the change of the velocity of the Earth during the year), in the case (b) the position of the source on the ellipse will be symmetrical to the position of any other ordinary source on its aberration ellipse (in the same neighborhood), moving the two in opposite directions. In this way, in the case (b) we will observe a motion of the source against the stream of all other (ordinary) sources on the sky, during a sufficient interval of time, of the order of some weeks. No such an effect occurs in case (a). (Fig. 5).

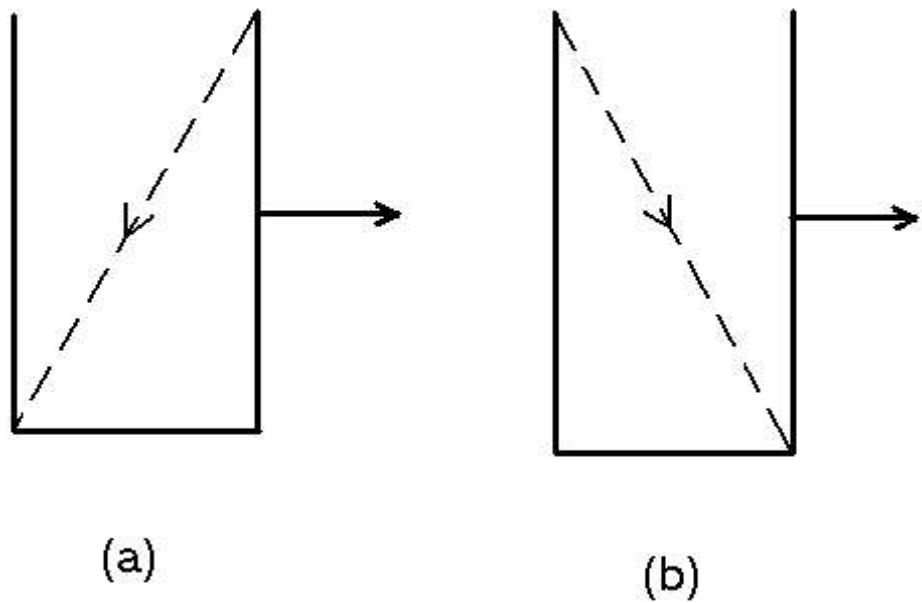

Fig. 5

Fig. 5. If we assume *positive* the aberration of normal light (a), the aberration of light moving backwards in time must be *negative* (b). The case corresponds to the telescope, moving with the velocity the arrows show, and the dashed line shows the path of light inside the telescope and with respect to it.

**Appendix A**

We have

$$\left.\begin{array}{ll} D = z^2 + 2z & \text{(a)} \\ D' = -z'^2 - 2z' & \text{(b)} \end{array}\right\} \quad (A1)$$

Setting $D = D' \equiv x$, we obtain the eqns.

$$\left.\begin{array}{ll} z^2 + 2z - x = 0 & (1) \\ -z'^2 - 2z' - x = 0 & (2) \end{array}\right\} \quad (A2)$$

Solving them, we get

$$\left.\begin{array}{ll} (1) \Rightarrow z = -1 + \sqrt{1+x} & \text{(a')} \\ (2) \Rightarrow z' = -1 + \sqrt{1-x}, & \text{(b')} \end{array}\right\} \quad (A3)$$

so that

$$\left.\begin{array}{ll} 1 + z = \sqrt{1+x} & (1') \\ 1 + z' = \sqrt{1-x} & (2') \end{array}\right\} \quad (A4)$$

We will show that $|z'| < z$, or, equivalently, that $z + z' > 0$, is false. In fact, from (1′) & (2′) we take

$$2 + (z + z') = \sqrt{1+x} + \sqrt{1-x}. \quad (A5)$$

Also, iff the above inequalities hold, then

$$[2 + (z + z')]^2 > 4. \quad (A6)$$

Now, from (A5), we have

$$[2 + (z + z)']^2 = 2 + 2\sqrt{(1+x)(1-x)}, \quad (A7)$$

or

$$[2 + (z + z')]^2 - 2 = 2\sqrt{(1+x)(1-x)}. \quad (A8)$$

Then, setting

$$[2+(z+z')]^2 - 2 \equiv X, \tag{A8´}$$

the inequality (A6) is equivalent to

$$X > 2, \tag{A9}$$

or

$$X^2 > 4. \tag{A10}$$

But, from (A8),

$$X^2 = 4(1+x)(1-x), \tag{A11}$$

or

$$X^2 = 4(1-x^2) < 4, \tag{A12}$$

QED.

**Appendix B**

Since, as we saw, our model operates in flat space-time, we must be able to extract (3.38) & (3.40) considering the redshift as Doppler shift. In fact, our Universe can be represented by the interior of the "light cone" at the origin O (see Fig.6). Suppose that an "antiparticle" of the cosmic fluid follows the world line AO, and when it is at A, at time t and at a distance D´ from the t-axis, it emits an antiphoton which reaches us (the observer) at time $t_0$, when it is really at A´, at a distance x from us. From geometrical considerations (similarity of triangles) we get from Fig.6 the relation

$$\frac{x}{D'} = \frac{t_0}{t}. \tag{B1}$$

Now, because of the flatness of space-time, the distances x and D´ can be taken either as luminosity distances or light distances – it is the same. Thus we may set $x = vt_0$ (v is the velocity of the "antiparticle") and $D' = c(t - t_0)$. Inserting these values of x & D' in (B1), we get

$$\frac{v}{c} = \frac{t - t_0}{t}, \tag{B2}$$

or

$$\frac{v}{c} = 1 - \frac{H}{H_0}. \tag{B3}$$

If we use the Hubble law
$$v = HD', \tag{B4}$$

substituting H from (B4) into (B3), we obtain

$$\frac{v}{c} = 1 - \frac{v/c}{H_0 D'/c}, \tag{B5}$$

so that

$$\frac{H_0 D'}{c} = \frac{v/c}{1 - v/c}. \tag{B6}$$

From the Doppler effect we have [8]

$$1 + z = \sqrt{\frac{1 + v/c}{1 - v/c}}, \tag{B6'}$$

so that

$$\frac{v}{c} = \frac{(1+z)^2 - 1}{(1+z)^2 + 1}. \tag{B7}$$

But in our case we have to take (in the antiworld)

$$\frac{v}{c} = \frac{(1+z')^2 - 1}{(1+z')^2 + 1}. \tag{B8}$$

Thus, substituting (B8) into (B6), and changing sign as we go to the world, we obtain

$$\frac{H_0 D'}{c} = -\frac{\dfrac{(1+z')^2 - 1}{(1+z')^2 + 1}}{\dfrac{(1+z')^2 + 1}{(1+z')^2 + 1} - \dfrac{(1+z')^2 - 1}{(1+z')^2 + 1}}, \tag{B9}$$

which finally gives

$$\frac{H_0 D'}{c} = \frac{-2z' - z'^2}{2}, \tag{B10}$$

or

$$D' = -\frac{cz'}{2H_0}(2+z'), \qquad (B11)$$

QED (cf. to (3.40)).

Concerning the world (light coming from the past), we similarly (but simpler) can get

$$D = \frac{cz}{2H_0}(2+z) \qquad (B12)$$

(cf. to (3.38).

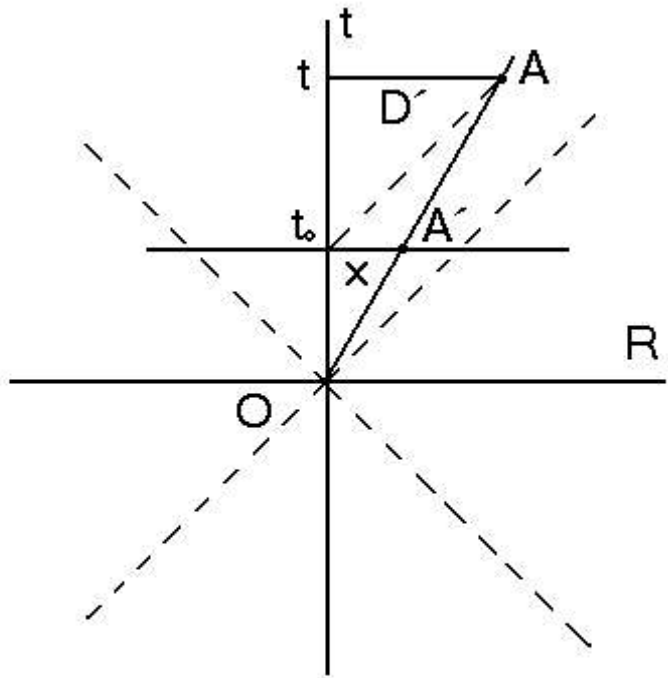

Fig. 6